\documentstyle[12pt]{article}
  \topmargin - 0.5 cm 
\newcommand{\ben}{\begin{enumerate}}
\newcommand{\een}{\end{enumerate}}
\newcommand{\be}{\begin{equation}}
\newcommand{\ee}{\end{equation}}
\newcommand{\bse}{\begin{subequation}}
\newcommand{\ese}{\end{subequation}}
\newcommand{\bea}{\begin{eqnarray}}
\newcommand{\eea}{\end{eqnarray}}
\newcommand{\bc}{\begin{center}}
\newcommand{\ec}{\end{center}}

\begin{document}
  \begin{titlepage}
  \scrollmode

\begin{center}
{\Large\bf A fractional supersymmetric oscillator} 

\vspace{0.5cm}
{\Large\bf and its coherent states}
\end{center}

\vspace{0.5cm}

\begin{center}
{\bf M. DAOUD$^1$ and M.~KIBLER$^2$}
\end{center}

\begin{center}
{$^1$D\'epartement de Physique}\\
{Facult\'e des Sciences}\\
{Universit\'e Ibnou Zohr}\\
{B.P. 28/S, Agadir, Morocco}
\end{center}

\begin{center}
{$^2$Institut de Physique Nucl\'eaire de Lyon}\\
{IN2P3-CNRS et Universit\'e Claude Bernard}\\
{43 boulevard du 11 novembre 1918}\\
{69622 Villeurbanne Cedex, France}
\end{center}

\vspace{2cm}
\begin{abstract}
We review some basic elements on $k$-fermions, which are objects 
interpolating between bosons and fermions. In particular, we
define $k$-fermionic coherent states and study some of their 
properties. The decomposition of a $Q$-uon into a boson and 
a  $k$-fermion  leads  to   a  definition    of   fractional 
supercoherent states.  Such states involve bosonic coherent
states and $k$-fermionic coherent states.  We construct an 
Hamiltonian which generalizes the ordinary (or $Z_2$-graded) 
supersymmetric oscillator Hamiltonian. Our Hamiltonian describes 
a  fractional (or $Z_k$-graded) supersymmetric oscillator  for
which the fractional supercoherent states are coherent states.
\end{abstract}

\vskip 5.65 true cm
\noindent 
Paper written from a plenary talk presented (by M.K.) at the  
{\bf Sixth International Wigner Symposium} 
(Istanbul, Turkey, 16 - 22 August 1999). Submitted for 
publication in Turkish Journal of Physics.  

\vfill
\thispagestyle{empty}
\end{titlepage}

\newpage

\section{Introduction}
In the last ten years, a considerable attention
has been paid to fractional supersymmetric 
(and para-supersymmetric) quantum mechanics [1-17]. 
Fractional supersymmetry corresponds to a  
$Z_k$-grading and can be realized in terms of
generalized Grassmann variables [1,3,18,19] or, in the
spirit of $q$-deformations of the oscillator 
algebra [20-24], in terms of para-fermionic variables [1-17] 
or $k$-fermionic variables [25,26]. On the other hand, 
$q$-deformed Glauber coherent states have been 
studied in several physical contexts [27-31].  

It is the aim of this paper to construct 
the Hamiltonian for one of the simplest 
fractional supersymmetric system, viz., the
$Z_k$-graded supersymmetric oscillator. The
construction of the Hamiltonian and its 
coherent states lies on the decomposition
of a $Q$-uon into a boson and a  $k$-fermion.

\section{On $k$-Fermionic Operators}
\subsection{The $k$-fermionic algebra $F_k$}
Let us start by defining what we shall refer to a $k$-fermionic algebra $F_k$.
The algebra $F_k$ is spanned by five operators $f_-$, 
                                               $f_+$, 
                                               $f_+^+$, 
                                               $f_-^+$
and $N$ through the following relations classified in three types.

\noindent (i) The $[f_-, f_+, N]$-type: 
$$
f_- f_+ - q f_+ f_- = 1
$$
$$
N f_-   -   f_- N   = - f_-,  \quad
N f_+   -   f_+ N   = + f_+
$$
$$
\left( f_- \right)^k   = 
\left( f_+ \right)^k   = 0
$$
\noindent (ii) The $[f_+^+, f_-^+, N]$-type:
$$
f_+^+ f_-^+ - {\bar q} f_-^+ f_+^+ = 1
$$
$$
N f_+^+     -          f_+^+ N     = - f_+^+,   \quad
N f_-^+     -          f_-^+ N     = + f_-^+
$$
$$
\left( f_+^+ \right)^k  = 
\left( f_-^+ \right)^k  = 0
$$
\noindent (iii) The $[f_-, f_+, f_+^+, f_-^+]$-type:
$$
f_- f_+^+ - q^{-{1 \over 2}} f_+^+ f_- = 0, \quad
f_+ f_-^+ - q^{+{1 \over 2}} f_-^+ f_+ = 0
$$
\noindent where the number 
$$
q := {\rm exp} \left( {2 \pi {\rm i} \over k} \right), 
\quad k \in {\bf N} \setminus \{ 0 , 1 \} 
$$
\noindent is a root of unity and ${\bar q}$ stands for the complex conjugate of $q$. The
couple $(f_-, f_+^+)$ of annihilation operators is connected to the couple 
       $(f_+, f_-^+)$ of creation     operators via the Hermitean conjugation relations
$$
f_+^+ = \left( f_+ \right) ^{\dagger}, \quad 
f_-^+ = \left( f_- \right) ^{\dagger}
$$
\noindent and $N$ is an Hermitean operator. It is clear that the case $k=2$ 
corresponds to fermions and the case $k \to \infty$ to bosons. In the two 
latter cases,  we can take $f_- \equiv f_+^+$ and $f_+ \equiv f_-^+$. In the 
other cases, the consideration of the two couples $(f_-, f_+^+)$ and 
$(f_+, f_-^+)$ is absolutely necessary. In the case 
where $k$ is arbitrary, we shall speak of $k$-fermions. 

\subsection{A representation of $F_k$}
A $k$-dimensional representation of $F_k$, on a $k$-dimensional Hilbert space 
spanned by the orthonormal set $\{ |n \rangle : n = 0, 1, \cdots, k-1 \}$, 
is easily obtained from 
\begin{eqnarray*}
f_- | n   \rangle &=& \left( \left[ n + s - {1 \over 2} \right]_q \right)^{{1 \over2}} 
    | n-1 \rangle \quad {\hbox{with}} \quad f_- |     0 \rangle = 0 \\
f_+ | n   \rangle &=& \left( \left[ n + s + {1 \over 2} \right]_q \right)^{{1 \over2}} 
    | n+1 \rangle \quad {\hbox{with}} \quad f_+ | k - 1 \rangle = 0 \\
f_+^+ | n \rangle &=& 
\left( \left[ n + s - {1 \over 2} \right]_{\bar q} \right)^{1 \over 2} | n-1 \rangle
                  \quad {\hbox{with}} \quad f_+^+ |   0 \rangle = 0 \\
f_-^+ | n \rangle &=& 
\left( \left[ n + s + {1 \over 2} \right]_{\bar q} \right)^{1 \over 2} | n+1 \rangle
                  \quad {\hbox{with}} \quad f_-^+ | k-1 \rangle = 0
\end{eqnarray*}
\noindent and 
$$
N | n \rangle = n | n \rangle
$$
\noindent The notation is as follows: we have $s := \frac{1}{2}$ and the factorials 
$[n]_{q} !$ and $[n]_{\bar q} !$ are defined by
$$
\lbrack n \rbrack_p ! := 
\lbrack 1 \rbrack_p 
\lbrack 2 \rbrack_p \cdots 
\lbrack n \rbrack_p \quad {\hbox{for}} \quad 
n \in {\bf N}^* \quad {\hbox{and}} \quad
\lbrack 0 \rbrack_p ! := 1
$$
\noindent with 
$$
[x]_p := {1 - p^x \over 1 - p} \quad {\hbox{for}} \quad x \in {\bf R}
$$
\noindent where $p = q, {\bar q}$. Note that the action of $f_-$, 
$f_+$, $f_+^+$ and $f_-^+$ on $| n \rangle$ is reminiscent of {\it finite} quantum mechanics 
discussed by many authors in various domains (for example, see [32] and [33]).  

\subsection{A Grassmanian realization of $F_k$}
It is possible to find a realization of the operators $f_-$, 
                                                      $f_+$, 
                                                      $f_+^+$ and
                                                      $f_-^+$ in terms of          
Grassmann variables $(\theta, {\bar \theta})$ and their $q$- and ${\bar q}$-derivatives
$(\partial_{\theta}, \partial_{\bar \theta})$. We take           Grassmann variables 
$\theta$ and ${\bar \theta}$ such that $ \theta^k = {\bar \theta}^k = 0 $ [1,3,18,19]. The sets
$\{ 1,       \theta , \cdots,       \theta  ^{k-1} \}$ and 
$\{ 1, {\bar \theta}, \cdots, {\bar \theta} ^{k-1} \}$ 
span the  same  Grassmann algebra $\Sigma_k$.  
The $q$- and ${\bar q}$-derivatives are 
formally defined by 
\begin{eqnarray*}
\partial_\theta f(\theta) &:=& {f(q\theta) - f(\theta) \over (q - 1) \theta} \\
\partial_{\bar \theta} g({\bar \theta}) &:=& 
{g({\bar q} {\bar \theta}) - g({\bar \theta}) \over ({\bar q} - 1) {\bar \theta}}
\end{eqnarray*}
\noindent Therefore, by taking 
$$
f_+   =       \theta ,   \quad  f_-   = \partial_      \theta,          \quad
f_-^+ = {\bar \theta},   \quad  f_+^+ = \partial_{\bar \theta}
$$
\noindent we have
$$
\partial_{\theta} \theta - q \theta \partial_{\theta} = 1,          \quad
\left( \partial_{\theta} \right)^k   =  \theta ^k     = 0
$$
$$
\partial_{\bar \theta} {\bar \theta} - {\bar q} {\bar \theta} \partial_{\bar \theta} = 1,        \quad
\left( \partial_{\bar \theta} \right)^k   =   {\bar \theta} ^k                       = 0
$$
$$ 
\partial_{\theta} \partial_{\bar \theta} - q^{-{1 \over 2}} \partial_{\bar \theta} \partial_{\theta} = 0, \quad
\theta {\bar \theta}                     - q^{+{1 \over 2}} {\bar \theta} \theta                     = 0
$$
\noindent Following Majid and Rodr\'\i guez-Plaza [19], we define the integration 
process
$$
\int d{      \theta}  \> {      \theta} ^n  = 
\int d{\bar {\theta}} \> {\bar {\theta}}^n := 0 \quad {\hbox{for}} \quad n = 0, 1, \cdots, k-2
$$
and
$$
\int d{\theta} \> {\theta}^{k-1} = \int d{\bar {\theta}} \> {\bar {\theta}}^{k-1} := 1
$$
\noindent which gives the Berezin integration for the particular case $k=2$. 

\section{Coherent States}
\subsection{The $k$-fermionic coherent states} 
We now define the states
$$
|\theta)        := \sum_{n=0}^{k-1} {       \theta ^n \over ([n]_q       !)^{1 \over 2} } 
\> | n \rangle
$$ 
\noindent and
$$
|{\bar \theta}) := \sum_{n=0}^{k-1} { {\bar \theta}^n \over ([n]_{\bar q}!)^{1 \over 2} }
\> | n \rangle
$$ 
\noindent as (finite) linear combinations of the eigenvectors $| n \rangle$ 
of the operator $N$. The states $| \theta )$ and $| {\bar \theta} )$
are $k$-fermionic coherent states in the sense that they satisfy the 
eigenvalue equations
$$
f_-   |       \theta  ) =       \theta  |       \theta  ), \quad
f_+^+ | {\bar \theta} ) = {\bar \theta} | {\bar \theta} )
$$
\noindent Similarly, we define
$$
( \theta | := \sum_{n=0}^{k-1} \> \langle n | \>
{ {\bar \theta}^n \over ([n]_{\bar q}!)^{1 \over 2} }
$$
\noindent and
$$
( {\bar \theta} | := \sum_{n=0}^{k-1} \> \langle n | \>
{ \theta^n \over ([n]_q!)^{1 \over 2} } 
$$
\noindent which are the dual states of the coherent states $| \theta )$ and 
$| {\bar \theta} )$, respectively. 

Note that, in the spirit of the works by Wang, Kuang and Zeng [31], 
it is also possible to define even and odd $k$-fermionic coherent states, 
which are eigenstates of $( f_- )^2$ and $( f_+^+ )^2$, 
and more generally
$m$-components of $k$-fermionic coherent states, which are eigenstates of 
$( f_- )^m$ and $( f_+^+ )^m$. 

We easily get the three following results.

{\bf Result 1}. We have
$$
(       \theta ' |      \theta  ) = {\rm e}_      q  ({\bar \theta}' \theta),  \quad  
( {\bar \theta}' |{\bar \theta} ) = {\rm e}_{\bar q} (\theta {\bar \theta}')
$$
\noindent where the $p$-deformed exponential ${\rm e}_p$ is defined by 
$$
{\rm e}_p : X \mapsto {\rm e}_p (X) := \sum_{n=0}^{k-1} \> { X^n \over [n]_p! }
$$
\noindent with $p= q, {\bar q}$. 

{\bf Result 2}. We have the over-completeness property
$$
\int \int d      \theta  \> |       \theta  ) \> 
\mu(\theta, {\bar \theta}) \> (       \theta  | \> d{\bar \theta} = 
\int \int d{\bar \theta} \> | {\bar \theta} ) \> 
\mu({\bar \theta}, \theta) \> ( {\bar \theta} | \> d      \theta  = 1
$$
\noindent where $\mu$ is defined via 
$$
\mu(\theta, {\bar \theta}) := \sum_{n=0}^{k-1} \> \left( [n]_q ! [n]_{\bar q} ! \right)^{1 \over 2} \>
\theta^{k-1-n} \> {\bar \theta}^{k-1-n}
$$
and the integrals have to be understood in terms of the above-mentioned
integration process.

{\bf Result 3}. By defining the coherence factor $g^{(m)}$, of order $m$, as
$$
g^{(m)} :=  { \left( \theta |  \left( f_-^+ \right)^m 
                               \left( f_-   \right)^m  | \theta \right) \over 
              \left( \theta |         f_-^+ 
                                      f_-              | \theta \right)^m }
$$
\noindent we obtain 
$$
\left| g^{(m)} \right| = \cases{ 0 \ {\rm for} \ m >    k-1 \cr \cr
                                 1 \ {\rm for} \ m \leq k-1
                               }
$$
\noindent and we conclude that a $k$-fermionic state cannot be occupied by more than 
$k-1$ identical $k$-fermions, a statement that induces a generalized Pauli exclusion 
principle.  

\subsection{On the $Q${\rm -uon} $\to$ {\rm boson} $+$ $k${\rm -fermion} decomposition} 
We know that a pair of $Q$-uons (with $Q$ generic) can give rise to a pair
of bosons and a pair of $q$-uons (with $q$ a root of unity) by making use 
of a limiting procedure where $Q \to q$ [34]. This is quite well-known in 
the case of the Macfarlane [21] (and 
Biedenharn [22]) $Q$-uons. The limiting procedure can be 
adapted to the case of the Arik and Coon [20]  $Q$-uons in the following way. 
We begin with a pair of $Q$-uons ($a_- , a_+$) satisfying  
$$
a_- a_+ - Q a_+ a_- = 1
$$
\noindent where $Q$ is generic. We assume that
$$
Q \to q := {\rm exp} \left( {2 \pi {\rm i} \over k} \right), 
\quad k \in {\bf N} \setminus \{ 0 , 1 \} 
$$
\noindent and then we take
$$
(a_{\pm})^k = 0 
$$
\noindent If we define
$$
b_{\pm} := \lim_{Q \to q} \frac{ (a_{\pm})^k }{ ([k]_Q!)^{\frac{1}{2}} }
$$
\noindent we get the result 
$$
b_- b_+ - b_+ b_- = 1
$$
\noindent so that the pair  ($b_- , b_+$) is a pair of ordinary bosons. We redefine
$a_{\pm}$ as 
$f_{\pm}$ for $Q= q$. Therefore, we also have a pair of
$k$-fermions ($f_- , f_+$) satisfying
$$
f_- f_+ - q f_+ f_- = 1 
$$
\noindent It can be proved that the $b$'s commute with the $f$'s.
As a conclusion, we have generated the set $\{ b_- , b_+ , f_- , f_+ \}$ from the set 
$\{ a_- , a_+ \}$. Indeed, the decomposition 
$\{ a_- , a_+ \} \to \{ b_- , b_+ , f_- , f_+ \}$ corresponds to the 
$Z$-line $\leftrightarrow$ $(z , \theta)$-superspace isomorphism 
described by Dunne {\it et al}.~[34] and Mansour {\it et al}.~[35].

\subsection{Fractional supercoherent states}
We may question: what happens to an ordinary $Q$-deformed coherent state 
$$
| Z ) := \sum_{n=0}^{\infty} \frac{ Z^n }{ ([n]_Q!)^{\frac{1}{2}} } \> | n \rangle
$$ 
\noindent (with $Q$ generic and $Z$ a complex number) when $Q$ goes to a root of unity~? By using 
the just described decomposition $Z \leftrightarrow (z , \theta)$, 
it is possible to show that the limit
$$
| z , {\theta} ) := \lim_{Z \to (z , \theta)} \> \lim_{Q \to q} \> | Z )
$$
\noindent can be written
$$
| z , {\theta} ) = | z ) \otimes | {\theta} )
$$
\noindent where 
$$
| z ) := \sum_{r = 0}^{\infty} \frac{ z^r }{ \sqrt{r!} } \> | r \rangle
$$ 
\noindent is an ordinary bosonic coherent state ($z$ is a 
bosonic complex variable) and 
$$
| {\theta} ) := \sum_{s=0}^{k-1} \frac{ \theta^s }{ ([s]_q!)^{\frac{1}{2}} } \> | s \rangle
$$ 
\noindent is a $k$-fermionic coherent state as introduced in subsection 3.1
(${\theta}$ is a $k$-fermionic Grassmann variable in $\Sigma_k$). 
The state 
$$
| z , {\theta} ) = \sum_{r=0}^{\infty} \frac{ z^r }{ \sqrt{r!} } \> | r \rangle 
           \otimes \sum_{s=0}^{k-1} \frac{ \theta^s }{ ([s]_q!)^{\frac{1}{2}} } \> | s \rangle
$$ 
shall be called a fractional supercoherent state. 

We note that the state $| z , {\theta} )$ is an eigenstate of the operator $b_- f_-$
with the eigenvalue $z \theta$. Furthermore, we can generate the state $| z , {\theta} )$
from the vacuum state 
$$
 |     0 \rangle \otimes |     0 \rangle \equiv 
 | r = 0 \rangle \otimes | s = 0 \rangle
$$ 
owing to  the  operator
$$
D_q (z , \theta) := \exp ( z b_+ )  \>  {\rm e}_q ( \theta f_+)
$$
\noindent As a matter of fact, we have
$$
| z , {\theta} ) = D_q (z , \theta) \> | 0 \rangle  \otimes  | 0 \rangle
$$
\noindent and thus the operator $D_q (z , \theta)$ plays the r\^ole of a 
displacement or dilation operator.

It is interesting to mention that,  for fixed $k$,  each fractional 
supercoherent state is a linear combination of the Vourdas [32]
coherent states on the Riemann surface $R_k = {\bf C}^*/{Z_k}$ with
coefficients in the Grassmann algebra $\Sigma_k$ spanned by 
$\{ 1, \theta, \cdots, \theta^{k-1} \}$. We note in passing that the
Vourdas coherent states
$$
| z, k, s ) := \sum_{r=0}^{\infty} \frac{ z^{kr} }{ \sqrt{r!} } \> | kr + s \rangle, \quad
        s = 0, 1, \cdots, k-1 
$$ 
correspond to a ${Z_k}$-grading of the Fock space associated to an 
ordinary harmonic oscillator. In terms of the latter states, the fractional 
supercoherent state $| z^k , {\theta} )$ can be developed,  modulo the correspondence 
$| kr + s \rangle \leftrightarrow | r \rangle \otimes | s \rangle$,  as 
$$
| z^k , {\theta} ) =               | z, k, 0   ) + 
                          \theta   | z, k, 1   ) + \cdots + 
\frac{ \theta^{k-1} }{ ([k-1]_q !)^{\frac{1}{2}} } | z, k, k-1 )
$$
with coefficients in $\Sigma_k$.

\section{A Fractional Supersymmetric Oscillator}
\subsection{Preliminaries}
At this stage, a legitimate question arises: what is the 
Hamiltonian having the fractional supercoherent states 
$| z , {\theta} )$ as coherent states~? An immediate 
answer can be obtained in the case $k = 2$. In this case,
we have
$$
| z , {\theta} ) = 
 \sum_{r=0}^{\infty} \frac{ z^r }{ \sqrt{r!} } \> | r \rangle \otimes | 0 \rangle + \theta
 \sum_{r=0}^{\infty} \frac{ z^r }{ \sqrt{r!} } \> | r \rangle \otimes | 1 \rangle
$$
which turns out to be a supercoherent state for an 
ordinary supersymmetric oscillator [36]. Such a supersymmetric 
oscillator corresponds to a $Z_2$-grading. Since the fractional 
supercoherent state $| z , {\theta} )$ corresponds to a $Z_k$-grading,
we foresee that the Hamiltonian we are looking for is the one for a
fractional supersymmetric oscillator corresponding to a $Z_k$-grading.
We now proceed to the construction of this Hamiltonian.  

\subsection{A $Z_k$-graded supersymmetric oscillator}
Our basic ingredients consist of a pair of ordinary bosons
$( b_- , b_+ )$ and a pair of $k$-fermions $( f_- , f_+ )$. The $f$'s satisfy
$q$-commutation relations and the $b$'s usual commutation relations 
(see above). In addition, the $f$'s commute with the $b$'s. Indeed, 
the pairs $( b_- , b_+ )$ and $( f_- , f_+ )$ may be considered as originating
from a pair of $Q$-uons $( a_- , a_+ )$ through the isomorphism 
between the braided line and the one-dimensional superspace. 

Let us define the operators $X_-$  and   $X_+$   by
\begin{eqnarray*}
X_- &:=& b_- \left[ f_-   +    \frac{(f_+)^{k-1}}{[k - 1]_q !} \right]       \\
X_+ &:=& b_+ \left[ f_-   +    \frac{(f_+)^{k-1}}{[k - 1]_q !} \right]^{k-1}
\end{eqnarray*}
\noindent and the Klein operator $K$ by
$$
K := f_- f_+ - f_+ f_-
$$
which reduces to the Witten operator for $k=2$. It is 
a simple matter of calculation to check that $X_-$, $X_+$ and $K$
satisfy
$$
X_- X_+ - X_+ X_- = 1
$$
$$
K X_+ - q X_+ K = 0, \quad K X_- - {\bar q} X_- K = 0
$$
$$
K^k = 1
$$
\noindent plus some ordinary commutation relations with the operator 
$M := X_+ X_-$, namely
$$
M X_-   -   X_- M   = - X_-,  \quad
M X_+   -   X_+ M   = + X_+
$$
$$
M K - KM = 0
$$
\noindent The operators $X_-$, $X_+$, $K$ and $M$ thus generate an extended Weyl-Heisenberg
algebra. We note that the form of the commutation relation  $[X_- , X_+] = 1$ is the same as
for the ordinary Weyl-Heisenberg algebra~; it   differs   from   the one used by
Plyushchay [12] and generalized by Quesne and Vansteenkiste [16]. 

The next step is to introduce the $k$ projection operators
$$
\Pi_{i} := \frac{1}{k}  \>  \sum_{s=0}^{k-1}  \>  q^{si} \> K^s, \quad i = 0, 1, \cdots, k-1
$$
\noindent for the cyclic group $Z_k$. We are thus in a position to define the two supercharges
$$
Q_- := X_- (1 - \Pi_{k-1}), \quad  Q_+ := X_+ (1 - \Pi_{0})
$$
\noindent among $k$ possible definitions. We easily verify that the $Q$'s satisfy the nilpotency 
relations
$$
\left( Q_- \right)^k = \left( Q_+ \right)^k = 0 
$$
\noindent Following the technique developed by Rubakov and Spiridonov  
in their work on para-fermions [1] (see also [6]), we introduce an Hamiltonian $H$ 
by means of the defining relation
$$
\left( Q_- \right)^{k-1} Q_+   +       \left( Q_- \right)^{k-2} Q_+ Q_- 
                               +   \cdots 
                               +   Q_+ \left( Q_- \right)^{k-1}
                               =       \left( Q_- \right)^{k-2} H
$$ 
\noindent This leads to the Hamiltonian
\begin{eqnarray*}
H = X_- X_+ \Pi_1 &+& \sum_{\ell = 2}^{k-1}      
                      (X_+ X_- - \ell + 1) (\Pi_0 + \Pi_1 + \cdots + \Pi_{k - \ell - 1 })   \\
                  &+& \sum_{\ell = 2}^{k-1} \ell (X_- X_+ + \frac{\ell - 1}{2}) \Pi_{\ell}
                                                + X_+ X_- (1 - \Pi_{k-1})
\end{eqnarray*}
\noindent which can be seen to satisfy the commutation relation
$$
H Q_{\pm} - Q_{\pm} H = 0
$$
and  thus  the two supercharges $Q_{-}$ and $Q_{+}$ can be regarded as constants of motion. 

\subsection{Some examples}
\subsubsection{Example 1}
As a first example, we take $k=2$, i.e., $q=-1$. Then, we have
\begin{eqnarray*}
X_- &:=& b_- \left( f_-   +    f_+ \right)       \\
X_+ &:=& b_+ \left( f_-   +    f_+ \right)
\end{eqnarray*}
\noindent and
$$
K := f_- f_+ - f_+ f_-
$$
\noindent where $( b_- , b_+ )$ are ordinary bosons 
and $( f_- , f_+ )$ ordinary fermions. The operators 
$X_-$, $X_+$ and $K$ satisfy
$$
X_- X_+ - X_+ X_- = 1
$$
$$
K X_+   +   X_+ K = 0, \quad 
K X_-   +   X_- K = 0
$$
$$
K^2 = 1
$$
\noindent which reflect bosonic  and  fermionic 
degrees of freedom. The projection operators
\begin{eqnarray*}
\Pi_0 &:=& \frac{1}{2} (1 + K) \\
\Pi_1 &:=& \frac{1}{2} (1 - K)
\end{eqnarray*}
\noindent are here simple chirality operators and the supercharges
\begin{eqnarray*}
Q_- &:=& X_- \Pi_0 \\
Q_+ &:=& X_+ \Pi_1
\end{eqnarray*}  
\noindent have the property
$$
\left( Q_- \right)^2 = \left( Q_+ \right)^2 = 0 
$$
\noindent The Hamiltonian $H$ assumes the form
$$
H = X_+ X_- \Pi_0   +   X_- X_+ \Pi_1 
$$
\noindent which can be rewritten as
$$
H = Q_- Q_+  +   Q_+ Q_-
$$
\noindent It is clear that $H$ commutes with $ Q_- $ 
and   $Q_+$.  In terms of boson and fermion operators, we have
\begin{eqnarray*}
Q_- &=& f_+ b_-   \\
Q_+ &=& f_- b_+
\end{eqnarray*}
\noindent and
$$
H = b_+ b_-    +   f_+ f_-
$$
\noindent so that $H$ corresponds to the ordinary (or $Z_2$-graded) 
supersymmetric oscillator whose energy spectrum $E$ is
$$
E = 1 \oplus 2 \oplus 2 \oplus \cdots
$$ 
\noindent with equally spaced levels, the ground state  being  a 
singlet (denoted by 1) and all the excited states being doublets
(denoted by 2). Finally, note that the fractional supercoherent state
$ | z , \theta ) $ with $k=2$ is a coherent state for the
Hamiltonian $H$ [36]. 

\subsubsection{Example 2}
We continue with $k=3$, i.e., 
$$
q = \exp \left( \frac{2 \pi {\rm i}}{3} \right)
$$
In this case, we take
\begin{eqnarray*}
X_- &:=& b_- \left[ f_-   +    \frac{(f_+)^2}{[2]_q!} \right]       \\
X_+ &:=& b_+ \left[ f_-   +    \frac{(f_+)^2}{[2]_q!} \right]^2
\end{eqnarray*}
\noindent and
$$
K := f_- f_+ - f_+ f_-
$$
\noindent where $( b_- , b_+ )$ are ordinary bosons 
and $( f_- , f_+ )$ are  $3$-fermions. We hence have
$$
X_- X_+ - X_+ X_- = 1
$$
$$
K X_+   -          q    X_+ K = 0, \quad 
K X_-   - \frac{1}{q}   X_- K = 0
$$
$$
K^3 = 1
$$
\noindent Our general definitions can be specialized to
\begin{eqnarray*}
\Pi_0 &:=& \frac{1}{3} (1 + q^3 K + q^3 K^2) \\
\Pi_1 &:=& \frac{1}{3} (1 + q^1 K + q^2 K^2) \\
\Pi_2 &:=& \frac{1}{3} (1 + q^2 K + q^1 K^2)
\end{eqnarray*}
\noindent for the projection operators and to  
\begin{eqnarray*}
Q_- &:=& X_- (\Pi_0  + \Pi_1) \\
Q_+ &:=& X_+ (\Pi_1  + \Pi_2) 
\end{eqnarray*}  
\noindent for the supercharges with the property
$$
\left( Q_- \right)^3 = \left( Q_+ \right)^3 = 0 
$$
\noindent By introducing the Hamiltonian $H$ via
$$
\left( Q_- \right)^{2} Q_+     +   Q_- Q_+ Q_- 
                               +   Q_+ \left( Q_- \right)^{2}
                               =   Q_- H
$$
\noindent we obtain
$$
H = \left( 2 X_+ X_-  -  1 \right) \Pi_0 +
    \left( 2 X_+ X_-  +  1 \right) \Pi_1 +
    \left( 2 X_+ X_-  +  3 \right) \Pi_2
$$
\noindent We can check that $H$ commutes with $ Q_- $ 
and   $Q_+$.  
\noindent The energy spectrum of $H$ reads 
$$
E = 1 \oplus 2 \oplus 3 \oplus 3 \oplus \cdots
$$ 
\noindent It contains equally spaced levels with a 
nondegenerate ground state (denoted as 1), a doubly 
degenerate first excited state (denoted as 2) and a
sequel of triply degenerate excited states (denoted as 3). 
  
\subsection{Concluding Remarks}
The work presented in this talk constitutes a further step
towards  fractional  supersymmetric  (or  para-supersymmetric) 
quantum mechanics with $N=2$ supercharges. We have given a realization
of an extended Weyl-Heisenberg algebra in terms of $k$-fermionic  
and purely bosonic operators. This algebra may be considered as 
originating from  the  decomposition of a $Q$-uon algebra 
into a $k$-fermionic algebra and a purely bosonic algebra.  As
a basic paradigm, we have investigated a fractional (or $Z_k$-graded) 
supersymmetric oscillator associated to our version of the Weyl-Heisenberg 
algebra. 

The $Z_k$-graded oscillator  described  in the present paper share some properties with 
the $C_{\lambda}$-extended harmonic oscillator introduced by Quesne and Vansteenkiste and
discussed at this symposium [16]. The main difference between the work 
in Ref.~[16] and our work lies in the fact that the  Weyl-Heisenberg algebra 
used by Quesne and Vansteenkiste involves the commutation relation
$$
X_- X_+ - X_+ X_- = \sum_{s=0}^{k-1} \> c_s K^s
$$
(where $c_s \in {\bf C}$ with $c_0 = 1$) instead of $X_- X_+ - X_+ X_- = 1$. 
The  Weyl-Heisenberg algebra considered in [16] generalizes the one used by 
Plyushchay [12] in its study of  the  Calogero-Vasiliev  
system and corresponding to $k=2$, viz.,
$$
X_- X_+ - X_+ X_- = 1 + c K
$$
(where $c \in {\bf C}$). In fact, it is also possible to find a realization
of the extended Weyl-Heisenberg algebra worked out by Plyushchay [12] 
and by Quesne and Vansteenkiste [16] in terms of $k$-fermionic operators
(or Grassmann variables) and bosonic operators (or complex variables)~; 
this will be the subject of a future work.  

Our approach is also concerned with the coherent states associated to
the $Z_k$-graded supersymmetric oscillator.  We have conjectured that these 
coherent states can be obtained from the decomposition of $Q$-deformed 
coherent states into $k$-fermionic coherent states and purely bosonic coherent 
states. On the basis of the results obtained in Ref.~[36], 
this conjecture is certainly true when $k=2$. In the case $k \ge 3$, 
we still have to prove it. In this respect, it would be 
useful as suggested by Solomon [37],  to examine the time evolution 
operator associated to the general Hamiltonian $H$ derived in this 
paper. Along this vein, it is a challenge to understand the connection
between $k$-fermions [25]  which  are of central importance in our 
approach and anyons  [38,39]  which  are also objects 
interpolating between fermions and bosons.    

The aspects of supersymmetry touched upon in this work
concern a small part of supersymmetry (namely, some aspects 
of supersymmetric quantum mechanics). Nevertheless, it may be
worth to close with a few words on the present status    
of supersymmetry in physics. From an optimistic point of view, 
supersymmetry is certainly a very appealing concept that proved
to be useful in mathematical physics and that is promising in 
particle physics. However, from a pessimistic  point of view, 
we have to realize that there is still no experimental evidence 
for supersymmetry except perhaps some hints in the so-called 
fractional Hall effect and in nuclear spectroscopy. (It is 
claimed from time to time that supersymmetry exists in atomic
nuclei, see for instance Ref.~[40]. In our opinion, supersymmetry 
in atomic nuclei simply relates the structure of odd-odd nuclei
to even-even and odd-$A$ systems rather than describes a 
symmetry between bosons and fermions.) No evidence for supersymmetry
exists in fundamental particle physics: no superparticle has been 
discovered yet and the limits on the Higgs mass afforded by 
supersymmetry have not led to the Higgs particle. The 
short-term hopes for the discovery of supersymmetry in
high energy physics are concentrated on analyses of the last e$^+$e$^-$ 
experiments of LEP at CERN  in 2000 and the forthcoming 
p${\bar {\rm p}}$ experiments of D\O~and~CDF at the Tevatron collider (Fermi-Lab). If nothing is 
found at CERN and Fermi-Lab, we shall have to wait the start 
of the pp experiments of LHC at CERN in 2005 (or more). We hope that supersymmetry
will not remain solely an elegant theoretical concept 
around the unification of internal and Lorentzian symmetries 
and that it 
will receive some experimental confirmation in a not too distant 
future.  

\medskip
\medskip
\smallskip
\smallskip

\noindent {\bf Acknowledgments} 

\medskip
\smallskip

\noindent One of the authors (M.K.) would like to thank 
the  Organizing   Committee   of
the Sixth International Wigner Symposium 
for inviting him to give a plenary talk.  The
authors are grateful to 
V.K.~Dobrev, J.P.~Gazeau, G.A.~Goldin, J.~Katriel, 
  A.~Mostafazadeh, C.~Quesne, A.I.~Solomon and A.~Vourdas  
for  comments, remarks  and   criticisms  on this work.

\end{document}